\documentstyle [12pt,axodraw] {article}

\catcode`@=12
\begin{document}
\thispagestyle{empty}
\begin{flushright} UCRHEP-T234\\July 1998\
\end{flushright}
\vspace{0.5in}
\begin{center}
{\Large	\bf Dynamical Neutrino Mass Matrix:\\
Large (Small) Mixing Solution to\\
Atmospheric (Solar) Neutrino Oscillations\\}
\vspace{1.5in}
{\bf Ernest Ma\\}
\vspace{0.3in}
{\sl Department of Physics, University of California\\}
{\sl Riverside, California 92521, USA\\}
\vspace{1.5in}
\end{center}
\begin{abstract}\
In the wake of the SuperKamiokande announcement of atmospheric neutrino 
oscillations which favor the large mixing of $\nu_\mu$ with $\nu_\tau$, 
there is considerable theoretical activity towards a possible understanding 
of the solar neutrino deficit also in terms of the mixing of $\nu_e$ 
with the other two neutrinos.  A specific dynamical model is presented here 
for the first time which naturally chooses the large (small) mixing solution 
to atmospheric (solar) neutrino oscillations.
\end{abstract}

\newpage
\baselineskip 24pt

Separate experimental evidences of neutrino oscillations at three (or four) 
very different wavelengths have been accumulating for many years\cite{1,2,3}. 
Recently the SuperKamiokande collaboration claimed\cite{4} the strongest 
signal to date of atmospheric neutrino oscillations which favor the large 
mixing of $\nu_\mu$ with $\nu_\tau$.  Together with other known constraints, 
a simple emerging conclusion is that there exists a neutrino mass eigenstate 
$\nu_\mu \cos \theta + \nu_\tau \sin \theta$ which is separated from the other 
states by a difference of mass squares ($\Delta m^2$) of about $10^{-3}$ to 
$10^{-2}$ eV$^2$ and $\sin^2 2 \theta$ is greater than about 0.8.  Within 
this context, if the solar neutrino deficit\cite{1} is also to be explained, 
we have basically only three options, as depicted in Fig.~1.

In Option (a), $\nu_e$ is the lowest lying state and mixes with $c \nu_\tau 
- s \nu_\mu$.  If their $\Delta m^2$ is from $10^{-6}$ to $10^{-5}$ eV$^2$, 
then the matter-enhanced solution\cite{5} to the solar neutrino deficit 
is possible.  If their $\Delta m^2$ is from $10^{-11}$ to $10^{-10}$ eV$^2$, 
then the vacuum-oscillation solution\cite{6} is possible.  In Option (b), 
$\nu_e$ is essentially degenerate with $c \nu_\tau - s \nu_\mu$, but it must 
still be below the latter to have the matter-enhanced solution.  On the other 
hand, the vacuum-oscillation solution does not depend on the sign of 
$\Delta m^2$ but requires near-maximal mixing between the two states to fit 
the data.  In Option (c), $\nu_e$ is placed below $\nu_\mu$ with $\Delta m^2$ 
of about 1 eV$^2$ to accommodate the LSND data\cite{3}.  In this case, a 
sterile neutrino $\nu_s$ is needed just above $\nu_e$ for the matter-enhanced 
solution to the solar neutrino deficit\cite{7,8}.  This scenario of course 
also works without $\nu_s$ if we were to discard the solar data.  It should 
also be mentioned that $\nu_e$ may still mix a little with $c \nu_\mu + s 
\nu_\tau$, as allowed by the SuperKamiokande experiment itself and within 
the constraint of the CHOOZ reactor data\cite{9}, as discussed already 
recently\cite{10,11}.

To gain some insight into how the above-mentioned three options may be 
realized theoretically, let us look at each in turn.  In Option (a), one 
mass is apparently much greater than the other two, whereas in Option (b), 
two nearly degenerate masses are much greater than the third.  Of course, 
in either scenario, it is also possible to add a common mass to all three 
states without changing their relative standings.  After all, neutrino 
oscillations only measure $\Delta m^2$ and not the individual masses. 
In Option (c), two nearly degenerate masses are much greater than two others. 
Note that whereas the gap between $\nu_e$ and $\nu_s$ is very small, the 
corresponding masses need not be degenerate in the sense that their ratio 
may in fact be very much less than unity.

If two neutrinos are nearly degenerate in mass, a natural explanation is 
that they are almost Dirac partners of each other.  In that case, they mix 
maximally.  In other words, there is a basis for which their $2 \times 2$ 
mass matrix is approximately of the form
\begin{equation}
{\cal M}_\nu \simeq \left( \begin{array} {c@{\quad}c} 0 & m \\ m & 0 
\end{array} \right).
\end{equation}
The mass eigenvalues of the above are $\pm m$ and the mixing is 45$^\circ$. 
The Majorana mass eigenstate which has the negative mass eigenvalue is 
redefined with a phase rotation, so that its mass becomes positive as 
required.  In Option (c), this would automatically set $c = s = 1/\sqrt 2$ 
and we have a theoretical understanding\cite{7} of why atmospheric neutrino 
oscillations favor $\sin^2 2 \theta = 1$.  In Option (b), $\nu_e$ would 
mix maximally with $c \nu_\tau - s \nu_\mu$ as proposed recently\cite{10}, 
and we have instead a theoretical understanding of why the vacuum-oscillation 
solution to the solar neutrino deficit favors $\sin^2 2 \theta = 1$.  In 
Option (a), it has also been assumed\cite{12} that $\nu_e$ mixes maximally 
with $c \nu_\tau - s \nu_\mu$ although the two are not necessarily 
degenerate in the sense that their mass difference may not be small compared 
to the larger of the two masses.  In both Options (a) and (b), the large 
mixing of $\nu_\mu$ with $\nu_\tau$, which has the maximum value\cite{13} 
of $\sin^2 2 \theta = 8/9$, is not explained without further theoretical 
assumptions.

Most models of the neutrino mass matrix simply assume a certain pattern or 
texture, using whatever arguments the authors find to be attractive.  A less 
arbitrary approach is to take the viewpoint that neutrino masses are in fact 
all zero at tree level, and ask how they can be generated dynamically in one 
loop.  The first and simplest such model was due to Zee\cite{14}.  The 
dynamics of this model automatically tell us that $\nu_\tau$ and a linear 
combination of $\nu_\mu$ and $\nu_e$ are almost Dirac partners.  If the 
latter is dominantly $\nu_\mu$, then we have maximal atmospheric neutrino 
oscillations and the LSND results are also explained\cite{15,16}.  To 
accommodate the solar data, $\nu_s$ has to be added [Option (c)], but if 
its mass is also generated dynamically, then a proper extension\cite{16} 
of the Zee model yields the interesting relationship
\begin{equation}
(\Delta m^2)_{\rm atm} \simeq 2 [(\Delta m^2)_{\rm solar} 
(\Delta m^2)_{\rm LSND}]^{1/2}
\end{equation}
which is well satisfied by the data.

In this paper I consider instead a natural dynamical realization of 
Option (a), where the mixing of $\nu_e$ with $c \nu_\tau - s \nu_\mu$ is 
small, for the matter-enhanced solution\cite{5} to the solar neutrino 
deficit.  The object is to obtain a large mass for $c \nu_\mu + s \nu_\tau$ 
and a small mass for $c \nu_\tau - s \nu_\mu$, with yet a smaller mass for 
$\nu_e$.  To do this, I supplement the Zee mechanism with a global $L_e$ 
symmetry, which is broken spontaneously as well as with explicit soft terms, 
by the interactions of the following new scalar particles.
\begin{equation}
(\eta^0, \eta^-), ~\chi^-, ~\zeta^0 ~: ~~~L_e = 1.
\end{equation}
Consider first the interactions given by
\begin{equation}
{\cal L}_1 = \sum_{i = \mu, \tau} f_i (\nu_i e - l_i \nu_e)\chi^+ + 
f'_i\bar e_R (\nu_i \eta^- - l_i \eta^0) + 
\mu_1 (\phi^- \eta^0 + \overline {\phi^0} \eta^-) \chi^+ + h.c.,
\end{equation}
where $(\phi^+, \phi^0)$ is the standard Higgs doublet with $L_e = 0$.  As 
shown in Fig.~2, the $2 \times 2$ mass matrix spanning $\nu_\mu$ and 
$\nu_\tau$ is now of the form
\begin{equation}
m_{ij} = {(f_i f'_j + f'_i f_j)\mu_1 v m_e \over 16 \pi^2 M^2},
\end{equation}
where $v \equiv \langle \phi^0 \rangle = 174$ GeV and $M$ is a large effective 
mass depending on $m_\eta^2$ and $m_\chi^2$.  Let $f_1/f_2 \equiv \tan \alpha$ 
and $f'_1/f'_2 \equiv \tan \beta$, then the above can be rewritten as
\begin{equation}
m_{ij} = m_0 \left[ \begin{array} {c@{\quad}c} 2 \cos \alpha \cos \beta & 
\cos \alpha \sin \beta + \sin \alpha \cos \beta \\ \cos \alpha \sin \beta + 
\sin \alpha \cos \beta & 2 \sin \alpha \sin \beta \end{array} \right].
\end{equation}
The trace of $m_{ij}$ is $2 m_0 \cos (\alpha - \beta)$ and its determinant is 
$-m_0^2 \sin^2 (\alpha - \beta)$.

Since $\tan \alpha$ and $\tan \beta$ are naturally of order unity, a very 
plausible scenario is to have $\cos (\alpha - \beta) \sim 1$ and 
$\sin^2 (\alpha - \beta) << 1$.  In that case, the mass eigenvalues are 
approximately $2 m_0$ and $-(m_0/2) \sin^2 (\alpha - \beta)$, corresponding 
to the eigenstates $c \nu_\mu + s \nu_\tau$ and $c \nu_\tau - s \nu_\mu$ 
respectively, where
\begin{equation}
{s \over c} \simeq {\sin (\alpha + \beta) \over 2 \cos \alpha \cos \beta},
\end{equation}
which is equal to one for $\alpha = \beta = 45^\circ$.  Numerically, let
$2 m_0 \simeq 0.07$ eV, then the other mass would be $2.3 \times 10^{-3}$ eV 
for $\sin^2 (\alpha - \beta) \simeq 0.13$.  These are very reasonable 
numbers and they are phenomenologically very relevant because they give 
$\Delta m^2 \simeq 5 \times 10^{-3}$ eV$^2$ for atmospheric neutrino 
oscillations and $\Delta m^2 \simeq 5 \times 10^{-6}$ eV$^2$ for solar 
neutrino oscillations.  To obtain $2 m_0 \simeq 0.07$ eV from Eq.~(5), we 
need $f f' \mu_1 / M^2 \simeq 6.2 \times 10^{-8}$ GeV$^{-1}$, which may be 
achieved with $f = f' = 0.05$, $\mu_1 = 100$ GeV, and $M = 2$ TeV.

So far, the one-loop diagram of Fig.~2 has yielded the following mass matrix 
in the basis ($\nu_e$, $c \nu_\tau - s \nu_\mu$, $c \nu_\mu + s \nu_\tau$):
\begin{equation}
{\cal M}_\nu = \left[ \begin{array} {c@{\quad}c@{\quad}c} 0 & 0 & 0 \\ 
0 & m_2 & 0 \\ 0 & 0 & m_3 \end{array} \right],
\end{equation}
where $m_2 \simeq 2.3 \times 10^{-3}$ eV and $m_3 \simeq 0.07$ eV.  To have 
matter-enhanced solar neutrino oscillations, $\nu_e$ must be made to mix with 
$c \nu_\tau - s \nu_\mu$ and its mass must be smaller than $m_2$.  To 
accomplish this, the neutral singlet $\zeta^0$ of Eq.~(3) is now used.  It 
contributes to the interactions given by
\begin{equation}
{\cal L}_2 = \mu_2 (\phi^+ \eta^- - \phi^0 \eta^0) \overline {\zeta^0} + 
\mu_3 (\phi^+ \phi^- + \phi^0 \overline {\phi^0}) \zeta^0 + \mu_4 (\eta^+ 
\eta^- + \overline {\eta^0} \eta^0) \zeta^0 + h.c.
\end{equation}
Whereas the $\mu_2$ term conserves $L_e$, the others do not.  Hence $L_e$ is 
softly broken explicitly.  Furthermore, as $\phi^0$ acquires a nonzero 
vacuum expectation value ($vev$), so must $\zeta^0$ and $\eta^0$.  Hence 
$L_e$ is also broken spontaneously, but there is no massless Goldstone boson 
because $L_e$ is already broken explicitly.  A nonzero $\langle \eta^0 
\rangle$ means that the charged-lepton mass matrix picks up nondiagonal terms 
from Eq.~(4).  The $3 \times 3$ matrix linking ($\bar e_L$, $\bar \mu_L$, 
$\bar \tau_L$) to ($e_R$, $\mu_R$, $\tau_R$) becomes
\begin{equation}
{\cal M}_l = \left[ \begin{array} {c@{\quad}c@{\quad}c} m_e & 0 & 0 \\ 
f'_\mu \langle \eta^0 \rangle & m_\mu & 0 \\ f'_\tau \langle \eta^0 \rangle 
& 0 & m_\tau \end{array} \right],
\end{equation}
and the rediagonalization of this matrix induces a mixing of $\nu_e$ with 
$\nu_\mu$ and $\nu_\tau$ of magnitude $f'_\mu \langle \eta^0 \rangle m_e / 
m_\mu^2$ and $f'_\tau \langle \eta^0 \rangle m_e / m_\tau^2$ respectively. 
The former should be much greater than the latter and it may not be 
negligible if $\langle \eta^0 \rangle$ is not small.  However, if $m_\eta^2$ 
is positive and large, then
\begin{equation}
\langle \eta^0 \rangle \simeq {\mu_2 v \langle \zeta^0 \rangle \over 
m_\eta^2}.
\end{equation}
In other words, a heavy $\eta$ may have a naturally small $vev$, as for 
example shown recently\cite{17}.

In the presence of a nonzero $\langle \eta^0 \rangle$, there is also a 
dynamical mass linking $\nu_e$ with $\nu_\mu$ and $\nu_\tau$ as shown in 
Fig.~(3), as well as one with itself as shown in Fig.~4.  The $3 \times 3$ 
matrix of Eq.~(8) is now changed:
\begin{equation}
{\cal M}_\nu = \left[ \begin{array} {c@{\quad}c@{\quad}c} m'' & c m' & s m' \\ 
c m' & m_2 & 0 \\ s m' & 0 & m_3 \end{array} \right],
\end{equation}
where
\begin{equation}
m' = {f_\tau m_\tau^2 \mu_1 \langle \eta^0 \rangle \over 16 \pi^2 v m_\chi^2},
~~~ m'' = {(f_\mu f'_\mu + f_\tau f'_\tau) m_e \mu_1 \langle \eta^0 \rangle^2 
\over 16 \pi^2 v m_\chi^2}.
\end{equation}
Let $\langle \eta^0 \rangle = 0.9$ MeV, and using the same values of 
$f = f' = 0.05$, $\mu_1 = 100$ GeV and $m_\chi = 2$ TeV as before, $m'$ is 
then $1.3 \times 10^{-4}$ eV and $m''$ is $1.9 \times 10^{-12}$ eV.  
Assuming $c \simeq 1/\sqrt 2$, the mixing of 
$\nu_e$ with $c \nu_\tau - s \nu_\mu$ in Eq.~(12) is
\begin{equation}
{c m' \over m_2} = 0.04,
\end{equation}
which yields the value of $\sin^2 2 \theta = 6 \times 10^{-3}$ for solar 
neutrino oscillations.

Going back to Eq.~(10), it is easy to check that the value of $\langle 
\eta^0 \rangle = 0.9$ MeV yields a mixing of $2 \times 10^{-6}$ for $\nu_e$ 
with $\nu_\mu$ which is clearly negligible.  From Eq.~(11), using again 
$m_\eta = 2$ TeV, it is seen that $\mu_2 \langle \zeta^0 \rangle = 21$ 
GeV$^2$, which is very reasonable.  The mass of the state dominated by 
$\nu_e$ is
\begin{equation}
m_1 = m'^2 / 2 m_2 = 3.7 \times 10^{-6} ~{\rm eV},
\end{equation}
the contribution $m''$ from Fig.~4 being negligible.

In summary, it has been shown in this paper for the first time how a 
specific dynamical model naturally chooses the large (small) mixing solution 
to atmospheric (solar) neutrino oscillations.  Unlike most other attempts 
to understand the present data, this is a complete theoretical model and 
not just an arbitrary speculation on the form of the neutrino mass matrix. 
The three approximate mass eigenstates are $\nu_\mu \cos \theta + \nu_\tau 
\sin \theta$, $\nu_\tau \cos \theta - \nu_\mu \sin \theta$ and $\nu_e$, 
with eigenvalues 0.07 eV, $2.3 \times 10^{-3}$ eV and $3.7 \times 10^{-6}$ eV 
respectively.  The $\nu_\mu - \nu_\tau$ mixing angle $\theta$ is naturally 
large and could be close to 45$^\circ$, thus accounting for the atmospheric 
data\cite{2,4}.  The mixing of $\nu_e$ with $\nu_\tau \cos \theta - \nu_\mu 
\sin \theta$ is about 0.04, resulting in a value of $\sin^2 2 \theta = 6 
\times 10^{-6}$, which is a suitable matter-enhanced solution to the solar 
data\cite{1}.

The $3 \times 3$ neutrino mass matrix, as given by Eq.~(12), 
is obtained dynamically in one loop, as depicted in Figs.~2 to 4.  This is 
achieved with a minimal set of new heavy scalar particles, as listed in 
Eq.~(3), which have $L_e = 1$, {\it i.e.} electron number just as $\nu_e$ 
and $e$.  In the limit of exact $L_e$ conservation, the mass matrix of 
Eq.~(8) is obtained.  This is sufficient to explain the atmospheric data. 
As $L_e$ is broken spontaneously (and with explicit soft terms so that a 
massless Goldstone boson will not appear), $\nu_e$ mixes with $\nu_\tau$ 
and matter-enhanced solar neutrino oscillations become possible.  This 
requires a small $\langle \eta^0 \rangle$, which is actually very 
natural\cite{17} for a positive and large $m_\eta^2$.  The above scenario 
[Option(a) of Fig.~1] is applicable as long as the LSND data\cite{3} are not 
considered.  If all positive neutrino-oscillation data are to be accommodated, 
an extension of the Zee model\cite{14} to include a sterile neutrino 
[Option(c)] could be the answer\cite{16}.  In either case, a dynamical 
explanation of the neutrino mass matrix works very well in the present 
experimental context.  

\newpage
\begin{center} {ACKNOWLEDGEMENT}
\end{center}

This work was supported in part by the U.~S.~Department of Energy under 
Grant No.~DE-FG03-94ER40837.

\bibliographystyle{unsrt}

\newpage
\begin{center}
\begin{picture}(480,240)(0,0)
\Line(40,200)(80,200)
\Text(85,200)[l]{$c \nu_\mu + s \nu_\tau$}
\Line(40,50)(80,50)
\Text(85,50)[l]{$c \nu_\tau - s \nu_\mu$}
\DashLine(40,25)(80,25)3
\Text(85,25)[l]{$\nu_e$}
\Text(80,-10)[c]{(a)}

\Line(200,200)(240,200)
\Text(245,200)[l]{$c \nu_\tau - s \nu_\mu$}
\DashLine(200,175)(240,175)3
\Text(245,175)[l]{$\nu_e$}
\Line(200,50)(240,50)
\Text(245,50)[l]{$c \nu_\mu + s \nu_\tau$}
\Text(240,-10)[c]{(b)}

\Line(360,200)(400,200)
\Text(405,200)[l]{$c \nu_\mu + s \nu_\tau$}
\Line(360,150)(400,150)
\Text(405,150)[l]{$c \nu_\tau - s \nu_\mu$}
\DashLine(360,35)(400,35)3
\Text(405,35)[l]{$\nu_s$}
\DashLine(360,25)(400,25)3
\Text(405,25)[l]{$\nu_e$}
\Text(400,-10)[c]{(c)}

\end{picture}
\vskip 0.8in
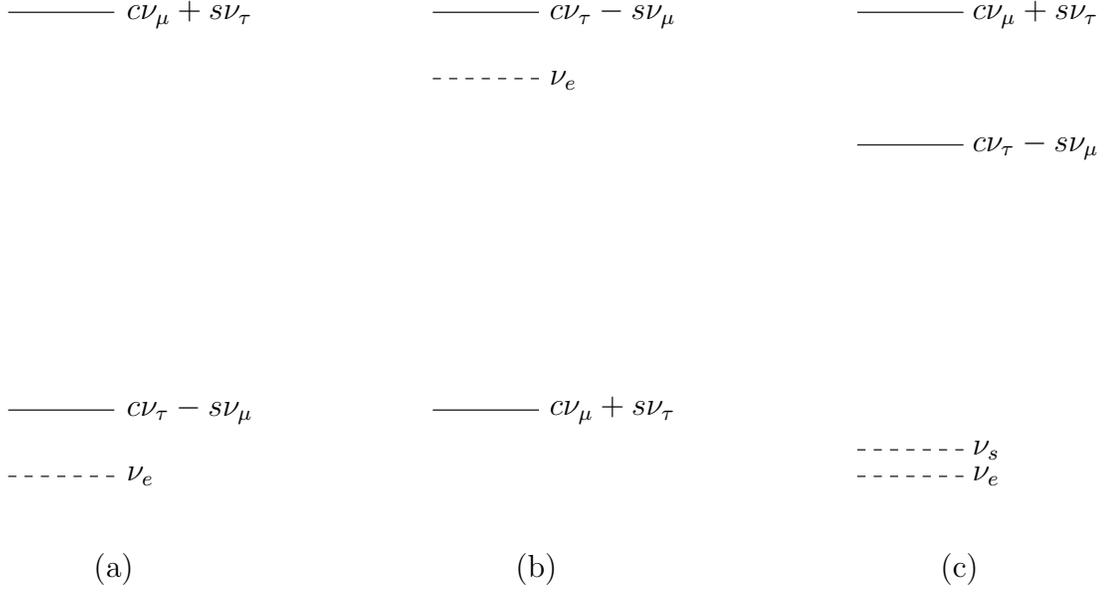
{\bf Fig.~1.} ~ Schematic drawings of possible neutrino mass hierarchies.
\vskip 0.8in

\begin{picture}(300,120)(0,0)
\ArrowLine(50,0)(90,0)
\ArrowLine(150,0)(90,0)
\ArrowLine(210,0)(150,0)
\ArrowLine(250,0)(210,0)
\DashArrowArcn(150,0)(60,180,90)3
\DashArrowArcn(150,0)(60,90,0)3
\DashArrowLine(150,95)(150,60)3
\Text(70,-8)[c]{$\nu_{\mu,\tau}$}
\Text(120,-8)[c]{$e_L$}
\Text(180,-8)[c]{$e_R$}
\Text(230,-8)[c]{$\nu_{\mu,\tau}$}
\Text(105,47)[r]{$\chi^-$}
\Text(199,47)[l]{$\eta^-$}
\Text(150,105)[c]{$\langle \phi^0 \rangle$}
\end{picture}
\vskip 0.5in
{\bf Fig.~2.} ~Dynamical mass generation for $\nu_\mu$ and $\nu_\tau$.
\end{center}

\newpage
\begin{center}
\begin{picture}(300,150)(0,0)
\ArrowLine(50,0)(90,0)
\ArrowLine(150,0)(90,0)
\ArrowLine(210,0)(150,0)
\ArrowLine(250,0)(210,0)
\DashArrowLine(150,60)(150,95)3
\DashArrowArcn(150,0)(60,180,90)3
\DashArrowArcn(150,0)(60,90,0)3
\Text(70,-8)[c]{$\nu_e$}
\Text(120,-8)[c]{$\tau_L$}
\Text(180,-8)[c]{$\tau_R$}
\Text(230,-8)[c]{$\nu_\tau$}
\Text(150,105)[c]{$\langle \eta^0 \rangle$}
\Text(105,47)[r]{$\chi^-$}
\Text(199,47)[l]{$\phi^-$}
\end{picture}
\vskip 0.5in
{\bf Fig.~3.} ~Dynamical mass linking $\nu_e$ with $\nu_\tau$.

\begin{picture}(300,150)(0,0)
\ArrowLine(50,0)(90,0)
\ArrowLine(150,0)(90,0)
\ArrowLine(210,0)(150,0)
\ArrowLine(250,0)(210,0)
\DashArrowLine(150,60)(150,95)3
\DashArrowLine(150,0)(150,-35)3
\DashArrowArcn(150,0)(60,180,90)3
\DashArrowArcn(150,0)(60,90,0)3
\Text(70,-8)[c]{$\nu_e$}
\Text(120,-8)[c]{$\mu_L,\tau_L$}
\Text(180,-8)[c]{$e_R$}
\Text(230,-8)[c]{$\nu_e$}
\Text(150,-43)[c]{$\langle \eta^0 \rangle$}
\Text(150,105)[c]{$\langle \eta^0 \rangle$}
\Text(105,47)[r]{$\chi^-$}
\Text(199,47)[l]{$\phi^-$}
\end{picture}
\vskip 1.0in
{\bf Fig.~4.} ~Dynamical mass for $\nu_e$.
\end{center}

\end{document}